\documentclass[pre,showpacs,byrevtex,amsmath,amssymb,preprint]{revtex4}

\usepackage{graphicx}
\usepackage{dcolumn}
\usepackage{bm}

\begin{document}
\preprint{}

\title[Short Title]{Relaxation back to equilibrium after cessation of shear for 
confined colloidal bilayers}

\author{Ren\'e Messina}
\affiliation
{Institut f\"ur Theoretische Physik II,
Heinrich-Heine-Universit\"at D\"usseldorf,
Universit\"atsstrasse 1,
D-40225 D\"usseldorf,
Germany}
\author{Hartmut L\"owen}
\affiliation
{Institut f\"ur Theoretische Physik II,
Heinrich-Heine-Universit\"at D\"usseldorf,
Universit\"atsstrasse 1,
D-40225 D\"usseldorf,
Germany}

\date{\today} 

\begin{abstract}
Crystalline bilayers of charged colloidal suspensions which are confined 
between two parallel plates and sheared via a relative motion of the 
two plates are studied by extensive Brownian dynamics computer simulations. 
The charge-stabilized suspension is modeled by a Yukawa pair potential. 
The unsheared equilibrium configuration are two crystalline layers with a 
nested quadratic in-plane structure.
For increasing shear rates $\dot \gamma$, we find the following
steady states: first, there is a static solid which is elastically sheared 
until a yield-stress limit is reached. Then there are two crystalline layers 
sliding on top of each other with a registration procedure. 
Higher shear rates melt the crystalline bilayers and even higher shear rates lead 
to a reentrant solid stratified in the shear direction.
This qualitative scenario is similar to that found in previous bulk simulations.
We have then studied the relaxation of the sheared steady state back to 
equilibrium after an instantaneous cessation of shear and found a nonmonotonic 
behavior of the typical relaxation time as a function of the shear rate $\dot \gamma$.
In particular, application of high shear rates accelerates the relaxation 
back to equilibrium since shear-ordering facilitates the growth of the equilibrium crystal. 
This mechanism can be used to grow defect-free colloidal crystals from strongly 
sheared suspensions. Our theoretical predictions can be verified in real-space 
experiments of strongly confined charged suspensions.
\end{abstract}

\maketitle

\section{Introduction}

A fundamental understanding of the different processes governing the
relaxation of metastable phases back to equilibrium is critical for
many basic questions in condensed matter physics and material
science. Also, relaxational processes are omnipresent in industrial
applications. Colloidal suspensions represent excellent model
systems where such questions can be studied directly in real-space
as the length scales are conveniently accessed experimentally, the
(variable) interactions can be described theoretically in a simple
way and the microscopic processes are rather slow as compared to
molecular materials. This has extensively exploited in previous
studies of interaction dependent equilibrium properties and dynamics
\cite{PuseyLH_D1,Müller_D1,Hartmut_Rev_D1,toprev}. One important example 
for a nonequilibrium steady state is a {\it sheared} colloidal suspension. 
It is known that application of shear may destroy the underlying equilibrium 
crystalline structure of the unsheared suspension \cite{sheardisorder} 
and can also lead to a reentrance ordering for high shear rates \cite{shearorder}. 
After cessation of shear the system will relax back to equilibrium from 
the sheared steady state. The microscopic details of this relaxation process 
are far from being resolved.

If an additional confinement between two parallel
plates is considered \cite{GrierMurray_D1}, various experiments
\cite{Wurth_D1,HSwetting_D1,Rheo_D1,Haw_D1,Dux_D1,Roufi_D1}
reveal a rich and subtle  influence of shear on the structure close to the wall.
Accordingly the relaxation back to equilibrium after cessation of shear 
is a fascinating but complex process which is a competition between 
wetting effects near the walls and bulk relaxation.
In experiments on strongly confined suspensions, for instance,
a complex pathway of the relaxation back to equilibrium was
obtained  \cite{Martensite_D1}: a bilayer bcc crystal was shear-molten to 
recrystallize as a buckled single layer triangular lattice which subsequently 
underwent a martensitic transition back to the equilibrium phase.

Most of the theoretical studies on  colloidal suspensions have addressed  
the influence of linear shear flow on the bulk structure via  non-equilibrium 
Brownian dynamics (NEBD) computer simulations \cite{Hess_D1} where hydrodynamic 
interactions are neglected and involve charged colloidal particles modeled by 
a Yukawa pair interaction 
\cite{Grest_D1,Stevens_PRL_1991,Stevens_D1,Chakrabarti_D1,Harrowell1_D1,NWagner1_D1,Cabane_D1,
NWagner2_D1,Blaak1,Blaak2}.
Shear-induced melting of colloidal bulk crystals and 
subsequent reentrant ordering at higher shear rates are confirmed by
simulation. However, simulational work including a wall acting on a 
sheared suspension is sparse; apart from a NEBD simulation in a channel
\cite{tube_D1} theoretical investigations were for a single
colloidal particle only \cite{sing1_D1,sing2_D1}.

In the present paper we  address the relaxation of
shear-induced structures after cessation of shear. We use
the standard Yukawa model for confined systems and employ
NEBD simulations. Here we focus on the
simple and transparent situation of a colloidal bilayers which are 
confined between two parallel plates and sheared via a relative motion of the two plates.
The reasons to do so is three-fold:
First, the equilibrium phase diagram for confined crystalline bilayers
interacting via a Yukawa pair potential is known from recent lattice-sum 
techniques at low temperatures \cite{Messina1}. This phase diagram
was recently confirmed in experiments on charged suspensions 
strongly confined between two glass plates \cite{Ana}. 
Second, the structure and the defects in a crystalline bilayer are easier to classify 
than in a multilayer. Last not least there are experimental studies for strongly confined 
situations which are not completely understood and are a challenge for a theoretical 
treatment \cite{Martensite_D1}. Recent simulation studies of Das and coworkers
\cite{Das1,Das2} have addressed similar questions regarding sliding bilayers.
The model employed in the studies of Das et al, however, is simpler than 
ours, it does not possess a  spatial dimension $z$ perpendicular to the plates
and hopping processes between the layers are ignored. 
Furthermore, the relaxation back to equilibrium is 
not investigated in Refs. \cite{Das1,Das2}.

In order to be specific, we chose the unsheared equilibrium
configuration to be  two crystalline layers with a nested quadratic 
in-plane structure.
This is the same staring configuration as used in the experiments 
\cite{Martensite_D1}.
For increasing shear rates $\dot \gamma$, we find the following
scenario of steady states: first, there is a static solid which is 
elastically sheared until
a yield-stress limit is reached. Then there are two crystalline layers 
sliding on top
of each other with a lock-in registration procedure similar to that 
observed in
recent experiments by Palberg and Biehl  \cite{Faraday,Biehl}. 
Higher shear rates melt the crystalline bilayers 
and even higher shear rates lead to a reentrant solid stratified in the 
shear direction.
This qualitative scenario is similar to that found in previous bulk simulations 
\cite{Grest_D1,Stevens_D1,NWagner1_D1}. The shear-induced ordering at
high shear rates  is reminiscent to the transition
towards lane formation in oppositely driven particles \cite{Dzubiella}.
We have then studied the relaxation of the sheared steady state back to 
equilibrium
after an instantaneous cessation of shear and found a nonmonotonic 
behavior of
the typical relaxation time as a function of the shear rate $\dot \gamma$.
In particular, application of high shear rates accelerates the relaxation 
back to equilibrium
via shear ordering in the steady state. This mechanism can be used to grow
defect-free colloidal crystals from strongly sheared suspensions as was 
proposed by Clark and coworkers \cite{Ackerson1,Ackerson2}. Our theoretical
predictions can be verified in experiments of confined charged suspensions
\cite{Martensite_D1,Faraday,Biehl}.

The paper is organized as follows: In section \ref{sec:ground_state}, 
we introduce the ground state model model for crystalline bilayers.
The nonequilibrium Brownian dynamics simulation technique is explained in 
section \ref{Sec.bd}.
Results are presented in section \ref{Sec.results}. 
Finally we conclude in section \ref{Sec.conclu}.

\section{The model}
\label{sec:ground_state}

In this part, we define our model. This is basically a generalization towards finite temperature
of the ground state model used in Ref. \cite{Messina1} concerning the {\it equilibrium} 
(i.e., without external applied shear flow) 
phase diagram of crystalline colloidal bilayers interacting via a Yukawa potential.
In detail, our system consists of two layers containing in total $N$
particles in the $(x,y)$ plane. The total area density of the two layers is
$\rho= N/A$ with $A$ denoting the layer area in the $(x,y)$ plane. 
The distance $D$ between the layers in the $z$ direction is prescribed by the external 
potential confining the system.
The  particles are interacting via the Yukawa pair potential 
%
\begin{equation}
\label{Eq.Yukawa}
V_{yuk}(r) =  V_0 \frac{\exp(-\kappa r)}{\kappa r},
\end{equation}
%
where $r$ is the center-center separation. 
The inverse screening length
$\kappa$  which governs the range of the interaction is given in terms
of the micro-ion concentration by Debye-H\"uckel
screening theory. The energy amplitude $V_0 = Z^2\exp (2\kappa R)\kappa/\epsilon
(1+\kappa R)^2$ scales with the square of the charges $Z$ of the particles
of physical hard core radius  $R$ 
reduced by the dielectric 
$\epsilon$ permittivity of the solvent ($\epsilon=1$ for the dusty plasma).
Typically, $Z$ is of the order of $100-100000$ elementary charges
such that $V_{yuk}(r)$ at typical interparticle distances can be much larger than the thermal 
energy $k_BT$ at room temperature
justifying formally zero-temperature calculations. Then the energy scale
is set by $V_0$ alone and  phase transitions in large bilayer systems
are  completely determined by 
two dimensionless parameters, namely the reduced layer density:
%
\begin{eqnarray}
\label{eq:eta}
\eta = \rho D^2 /2
\end{eqnarray}
%
and the reduced screening strength: 
%
\begin{eqnarray}
\label{eq:lambda}
\lambda = \kappa D
\end{eqnarray}
%
For zero temperature, the stable state is solid  but  different
crystalline structures of the bilayers are conceivable. 
The result for the phase diagram  in a $(\eta,\lambda)$-map
can be found in  Ref. \cite{Messina1}.
Here, we explore the same model for finite temperature by computer simulation. 
%

\section{The nonequilibrium Brownian dynamics computer simulation 
 \label{Sec.bd}}

\subsection{Simulation method}
\label{sec:simulation-method}

\begin{figure}
\includegraphics[width = 8.5 cm]{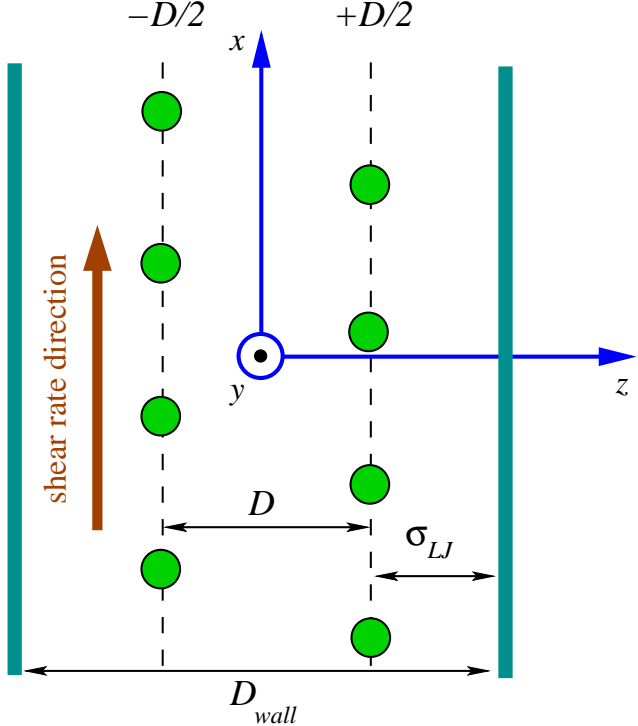}
\caption{View in the $(x,z)$ plane of the setup of the colloidal bilayer 
confined between two soft walls.
}
\label{Fig.setup} 
\end{figure}

Here, we provide a detailed description of our Brownian dynamics
method that was used to investigate {\it non-equilibrium} sheared
colloidal bilayers (at finite temperature).
A schematic setup of the system in the $(x,z)$ plane is
depicted in Fig. \ref{Fig.setup}.
The integration scheme for our model system in the
presence of an external steady shear rate $\dot \gamma$ reads:
%
\begin{eqnarray}
\label{eq.BD_eom}
{\bf r}_i(t + \delta t) = {\bf r}_i(t) 
+ \frac{D_0}{k_BT} {\bf F}_i(t) \delta t 
+ \delta {\bf W}_i 
+  \dot \gamma z_i(t) \delta t {\bf e}_x.  
\end{eqnarray}
%
Thereby $ {\bf r}_i(t)=[x_i(t),y_i(t),z_i(t)]$ is the position of the $i-$th
colloidal particle at time $t$ and $D_0$ denotes its free diffusion constant. 
All the contributions to the equation of motion \eqref{eq.BD_eom} 
are explained below.

Within a small time interval $\delta t$, that
particle moves under the influence of the sum of conservative forces 
${\bf F}_i(t)$ stemming from: 
(i) The pair interaction $V_{yuk}$ [see Eq. \eqref{Eq.Yukawa}] between particle $i$ and
    the neighboring ones.
(ii) The repulsive interaction with the soft wall(s) whose potential of interaction,
      $V_{wall}$, is modeled as follows
      %
      \begin{equation}
      \label{eq.BD_wall} 
      V_{wall} (z) = \left\{
      \begin{array}{ll}
      \displaystyle \alpha \epsilon_{LJ} \left[   
       \left( \frac{\sigma_{LJ}}{\frac{D_{wall}}{2}-|z|} \right)^{10} 
      -\left( \frac{\sigma_{LJ}}{\frac{D_{wall}}{2}-|z|} \right)^{4} 
      \right] + \epsilon_{LJ}, & 
      \textrm{for} \quad \displaystyle 
      \frac{\frac{D_{wall}}{2}-|z|}{\sigma_{LJ}} \geq  \left( \frac{5}{2} \right)^{1/6} , \\ \\
      0, \qquad \textrm{for} \quad \displaystyle 
      \frac{\frac{D_{wall}}{2}-|z|}{\sigma_{LJ}} < \left( \frac{5}{2} \right)^{1/6},
      \end{array}
      \right.
      \end{equation}
      %
where
$$
\alpha =
-\left[ \left( \frac{1}{z_{min}} \right)^{10} 
       -\left( \frac{1}{z_{min}} \right)^{4} 
\right]^{-1} = 3.07002 \dots
$$
[with $z_{min}=\left(\frac{5}{2}\right)^{1/6} \sigma_{LJ}$ minimizing
$V_{wall}$ in Eq. \eqref{eq.BD_wall}]
so that $V_{wall} (\sigma_{LJ})=\epsilon_{LJ}$ .
This (truncated and shifted) $10-4$ Lennard-Jones  potential given by Eq. \eqref{eq.BD_wall} 
assumes that we have thin soft walls .
Note that the use of a $9-3$ Lennard-Jones potential corresponding to
{\it semi-infinite} walls would not change qualitatively the results.
Also the use of charged hard walls would not affect our main results.

Furthermore due to the presence of the solvent, the particles experience
(i) a friction whose constant is given by $k_BT/D_0$ and (ii) 
random displacements, $\delta {\bf W}_i$. 
Those latter are sampled from a Gaussian distribution 
with zero mean and variance $2D_0 \delta t$ (for each Cartesian component).
The last term in Eq. \eqref{eq.BD_eom} represents the applied shear in the
$x-$direction, and imposes an explicit linear flow field.
The zero velocity plane of the imposed shear lies at the midplane between the plates.
%
%

\subsection{Parameters}
\label{sec:parameters}

The colloidal particles are confined in a rectangular $L \times L \times D_{wall}$
box where periodic boundary conditions are applied in the $(x,y)$ directions.
The system is made up of $N=800$ particles (i. e., 400 particles per layer). 
The units are set as follows: $k_BT=1/\beta$ sets the energy scale, 
the (typical average) interlayer separation $D=D_{wall}-2\sigma_{LJ}$ 
(see also Fig. \ref{Fig.setup}) sets the length scale, 
and $\tau=D^2/D_0$ sets the time scale. 
For the Yukawa interparticle interaction [see Eq. \eqref{Eq.Yukawa}] we 
choose $\beta V_0=6000$, whereas for the wall-particle interaction  
[see Eq. \eqref{eq.BD_wall}] we choose $\beta \epsilon_{LJ} = 1$ and  
$\sigma_{LJ} = 0.1 D$. The time step was set to $\delta t=10^{-5}\tau$.
The reduced colloidal particle density  is set to $\eta=\frac{ND^2}{2L^2}=0.24$ 
(so that $L=40.82D$) and the reduced screening is $\lambda=\kappa D=2.5$. 
Those latter parameters lead to the staggered square phase in the ground state
(or at very low temperature) as can be seen on the phase diagram from 
Ref. \cite{Messina1}.

\begin{figure}
\includegraphics[width = 8.5 cm]{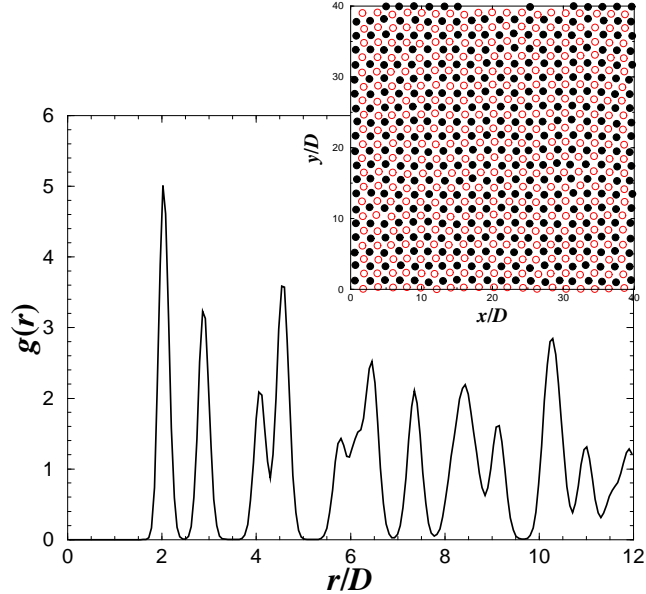}
\caption{Intralayer $(x,y)$ pair distribution function $g(r=\sqrt{x^2+y^2})$ 
at equilibrium ($\dot \gamma=0$). 
The inset shows a simulation snapshot where the filled (open) circles 
represent particles belonging to the upper (lower) layer.} 
\label{Fig.g_of_r_equil} 
\end{figure}

\begin{figure}
\includegraphics[width = 8.5 cm]{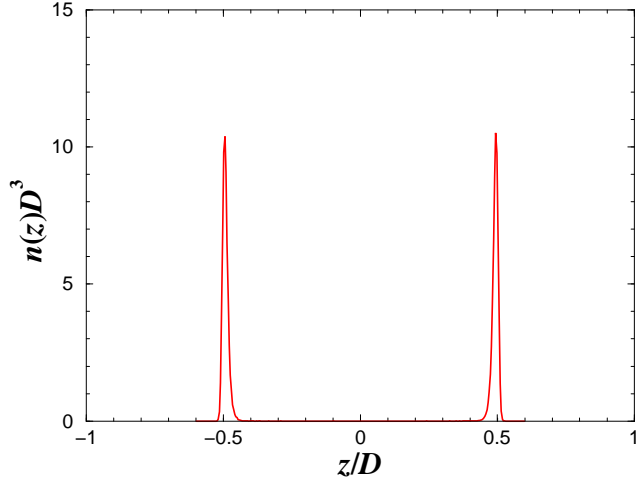}
\caption{Laterally averaged inhomogeneous particle 
density $n(z)$ at equilibrium ($\dot \gamma=0$). }
\label{Fig.density_equil} 
\end{figure}

The equilibrium (i.e., $\dot \gamma=0$) properties of our model system are 
obtained over a period of $10^6$ BD time steps (i. e., $10 \tau$.)
The corresponding in-plane $(x,y)$ pair distribution function $g(r)$ 
is shown in Fig. \ref{Fig.g_of_r_equil}.
It clearly shows a high degree of ordering as characterized by the
pronounced peaks and the deep minima. The snapshot also provided in 
Fig. \ref{Fig.g_of_r_equil} confirms the square lattice structure 
expected for those parameters.
To quantify the layer extension in the $z$-direction we have also
plotted the the particle density $n(z)$ that can be found on 
Fig. \ref{Fig.density_equil}.
The mean interlayer separation is then given by
$2\int_{0}^{D_{wall}/2} zn(z)L^2dz \approx 0.99D$, so that
(in practice) $D$ corresponds indeed to the interlayer separation.

%
%

%
%

%

\section{Results
 \label{Sec.results}}


\subsection{Effect of shear flow}
\label{sec:effect-shear-flow}

Starting from the equilibrium configuration described in the previous section, 
an external shear is applied during a period of $4 \times 10^6$ BD steps (i.e., $40 \tau$).
A steady is reached after typically $10 \tau$, and subsequent measurements
are performed over a typical period of 20 $\tau$.

%
\begin{figure}
\includegraphics[width = 14 cm]{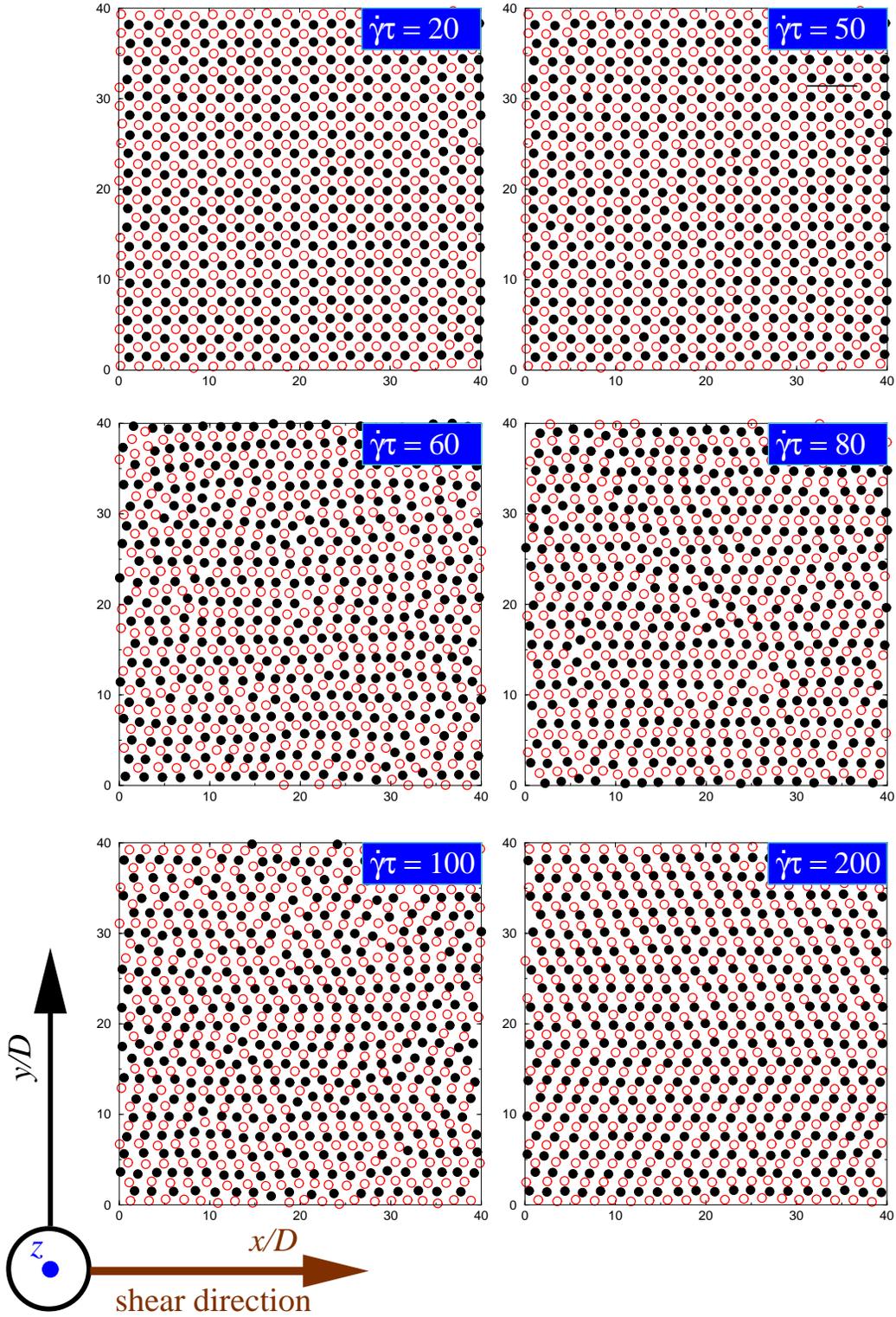}
\caption{Simulation snapshots for different values of the shear rate $\dot \gamma$
(as indicated) where the filled (open)  circles represent particles belonging to the upper 
(lower) layer.} 
\label{Fig.snap_shear} 
\end{figure}
%

It is instructive to start our study by analyzing the microstructures
reported in Fig. \ref{Fig.snap_shear} corresponding to different $\dot \gamma$.
From a structural point of view one can (qualitatively) identify three regimes: 
%
\begin{itemize}
\item 
At sufficiently low shear rates (here $\dot \gamma = 20/\tau$ and  
$\dot \gamma = 50/\tau$), it can be seen that the crystalline structure (namely square)
as well as the degree of ordering are conserved compared to the equilibrium situation 
(i.e., $\dot \gamma = 0$ - see Fig. \ref{Fig.g_of_r_equil}). 
Consequently we are in an elastic regime where the applied shear flow is
smaller than the {\it yield stress}.
\item 
For intermediate shear rates (here $\dot \gamma = 60/\tau$ and  
$\dot \gamma = 80/\tau$), there is a (relative) strong disorder
and the structure can therefore be qualified as liquid. In other words
we have to deal with a {\it shear induced melting}.  
\item
At high shear rates (here $\dot \gamma = 100/\tau$ and $\dot \gamma = 200/\tau$),
the system gets \textit{again ordered} (especially for the highest shear rate $\dot \gamma = 200/\tau$)
but exhibits a different (intralayer) crystalline symmetry (namely a triangular lattice) 
than the equilibrium one. Consequently, we have a reentrant behavior 
concerning the \textit{intralayer}-ordering upon shearing.
\end{itemize}
%

%
\begin{figure}
\includegraphics[width = 8.5 cm]{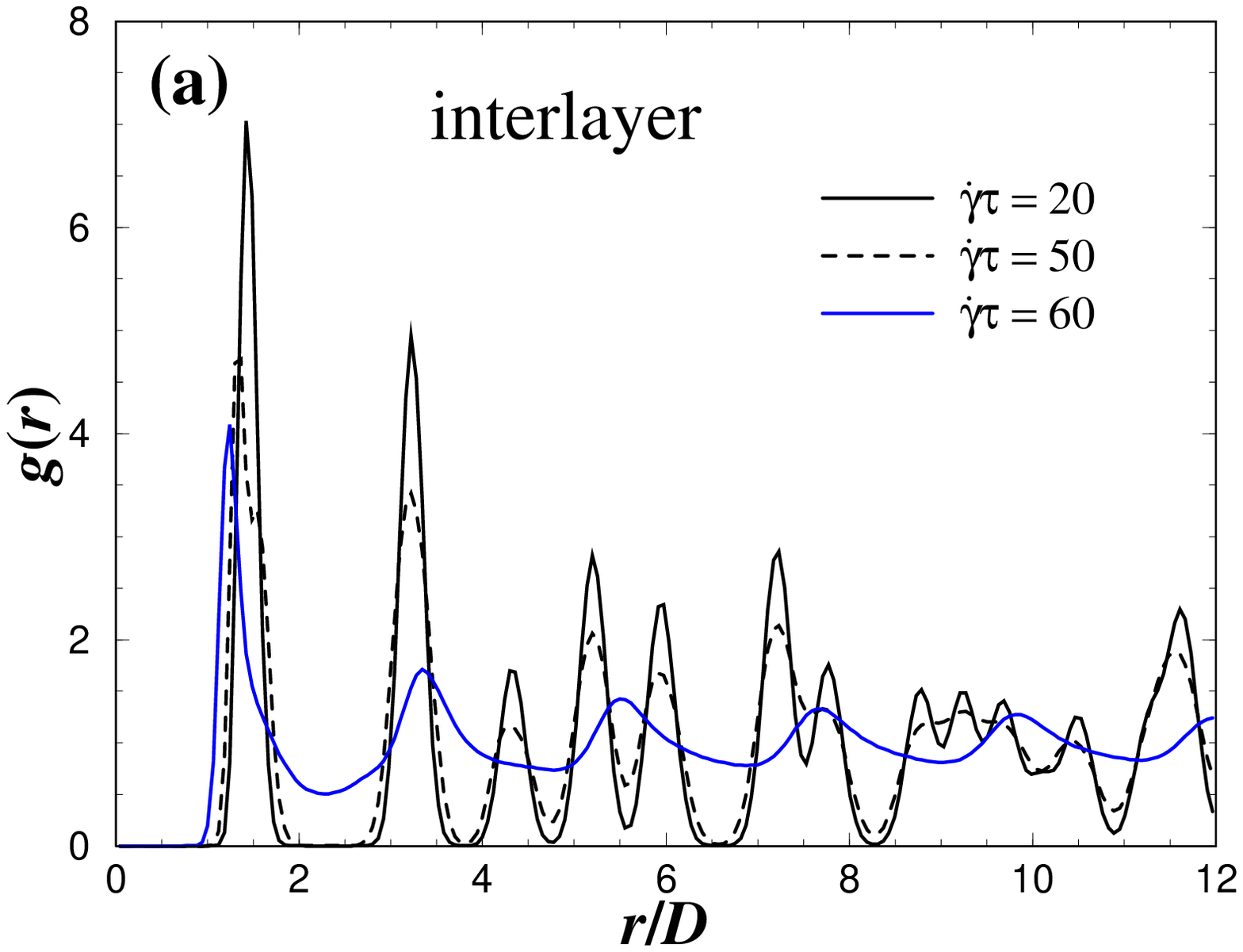}
\includegraphics[width = 8.5 cm]{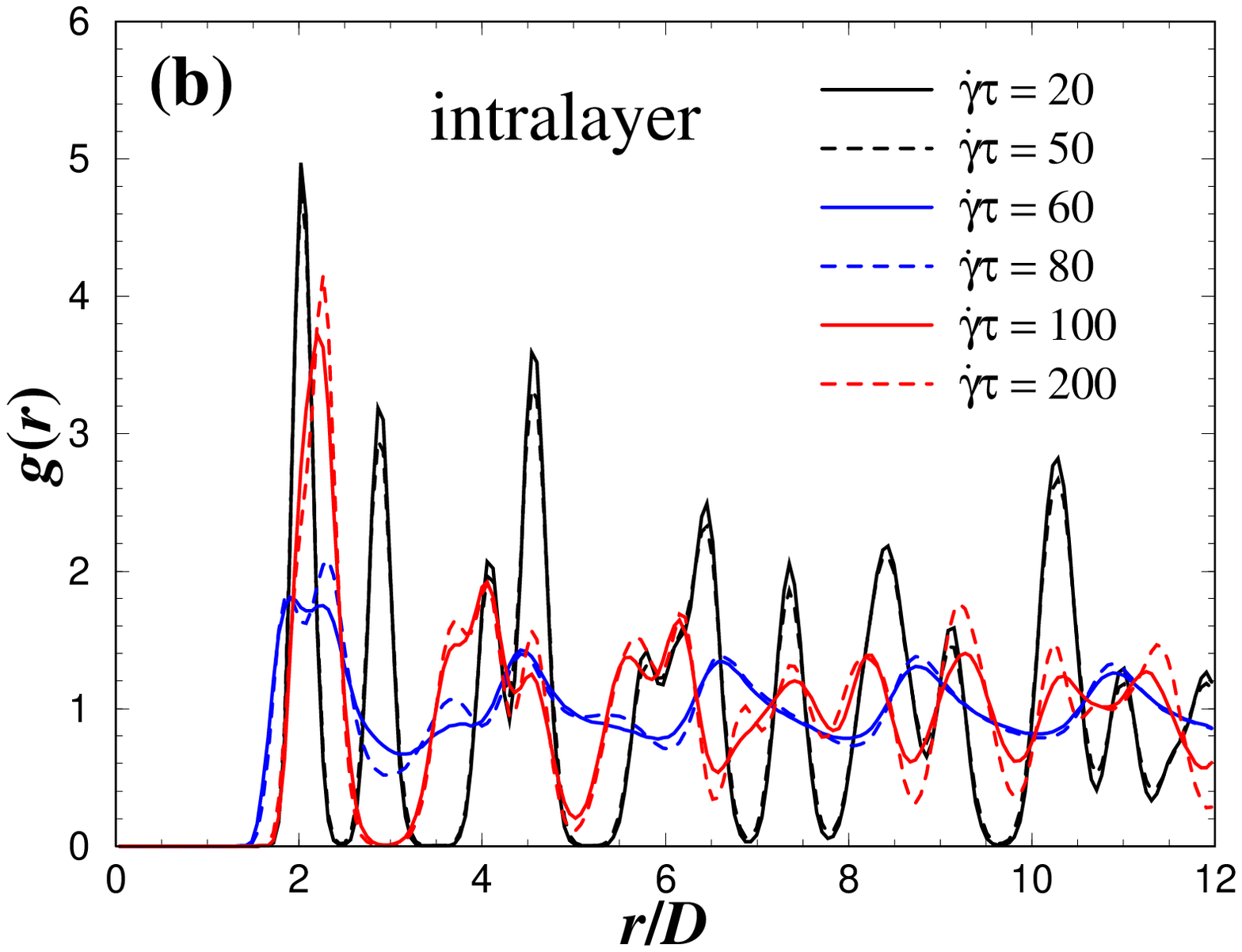}
\caption{
(a){\it Interlayer} $(x,y)$ pair distribution function $g(r=\sqrt{x^2+y^2})$ for
small values of $\dot \gamma$ (as indicated in the legend).
(b){\it Intralayer} $(x,y)$ pair distribution function $g(r=\sqrt{x^2+y^2})$ 
for different  values of $\dot \gamma$ (as indicated in the legend). 
The corresponding simulation snapshots are displayed in Fig. \ref{Fig.snap_shear}}.    
\label{Fig.g_of_r_shear} 
\end{figure}
%

In order to obtain a more quantitative description of these 
$\dot \gamma$-dependent structural properties,
we have also computed the (azimuthally averaged) {\it interlayer}- 
and {\it intralayer}-pair-distribution-functions 
$g(r=\sqrt{x^2+y^2})$ for different $\dot \gamma$.
The results are presented in Fig. \ref{Fig.g_of_r_shear}. 

The elastic behavior and the yield stress can be best understood
by considering the interlayer and intralayer $g(r)$. 
From Fig. \ref{Fig.g_of_r_shear}, we see that at weak shearing 
(here $\dot \gamma = 20/\tau$), the intralayer crystalline structure 
as well as the interlayer-lattice-correlation remain unchanged compared
to the $\dot \gamma = 0$ case (the latter is not reported in Fig. \ref{Fig.g_of_r_shear}).
At larger shear rate (here $\dot \gamma = 50/\tau$) the degree
of interlayer-lattice-correlation gets weaker than that of 
the intralayer one.
A closer look at Fig. \ref{Fig.g_of_r_shear}(a) reveals that, 
for $\dot \gamma = 50/\tau$, the first peak is (asymmetrically) splitted 
into two neighboring peaks. 
This is the signature of a small relative displacement of the two square layer-lattices.
Upon further increasing the shear rate (now at $\dot \gamma = 60/\tau$),
the bilayer becomes a liquid, demonstrating that there is a yield stress
$\dot \gamma_0$ whose value is such that $50/\tau < \dot \gamma_0 < 60/\tau$.

Above the yield stress, the intralayer $g(r)$ [see Fig. \ref{Fig.g_of_r_shear}(b)] 
exhibits a non-trivial behavior with respect to the applied shear flow, 
in agreement with our previous discussion on the microstructures depicted in 
Fig. \ref{Fig.snap_shear}.
More precisely, at intermediate values of $\dot \gamma$ (here $60/\tau$ and 
$80/\tau$), the intralayer layer structure corresponds to a liquid one.
Nonetheless and interestingly, at first neighbor separations, 
the square structure locally persists, but in coexistence with a triangular structure,  
as indicated by the broadened (splitted) first peak.
This feature can also be nicely visualized on the snapshots from Fig. \ref{Fig.snap_shear}. 
At high shear rates (here $100/\tau$ and $200/\tau$), there is a strong ({\it re})ordering
into a triangular lattice as indicated by the shifted first pronounced peak 
(especially for  $\dot \gamma=200/\tau$). However the degree of ordering reported for 
those highly sheared structures is not as high as that observed below yield stress.

To further quantify the behavior of highly sheared colloidal bilayers and also to provide 
a dynamical information, we are going to examine the (dimensionless) modified 
Lindemann parameter, $\Gamma_L(t)$, that is defined as follows
%
\begin{eqnarray}
\label{eq.lindemann}
\Gamma_L(t) & = &  \frac{\langle u^2(t)\rangle}{D^2}, 
\end{eqnarray}
%
where $\langle u^2(t) \rangle$ corresponds to the difference in the mean square displacement
of neighboring particles from their initial sites ${\bf r}_0 = {\bf r}(t=t_0)$. 
More explicitly, $\langle u^2(t)\rangle$ can be written as
%
\begin{eqnarray}
\label{eq.u2}
\langle u^2(t) \rangle & = & 
\left \langle 
\frac{1}{N}\sum_{i=1}^{N} \frac{1}{N_b} \sum_{j=1}^{N_b}
\left[({\bf r}_i(t) - {\bf r}_{i}(t_0)) - ({\bf r}_j(t) - {\bf r}_{j}(t_0))\right]^2
\right \rangle , 
\end{eqnarray}
%
where ${\bf r}_i(t)=[x_i(t),y_i(t)]$, $\langle \dots \rangle$ denotes an averaging over BD steps, the index $j$
stands for the $N_b$ nearest neighbors of particle $i$ lying in the same 
upper or lower layers. Typically, for a (local) triangular lattice 
environment $N_b=6$ while for a rectangular one $N_b=4$. 
Besides, we also average over several reference times $t_0$ to improve
the statistics. Due to the finite size of the simulation box, one is 
typically limited to observation times $\Delta t_{obs}$ of the order of 
$\Delta t_{obs} \approx L / {\dot \gamma_{max} D} \approx 0.2\tau$ 
(by taking here $\dot \gamma_{max}=200/\tau$).

\begin{figure}
\includegraphics[width = 8.5 cm]{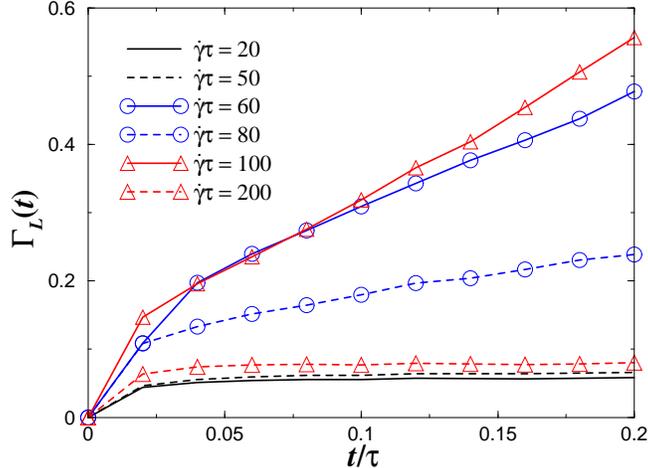}
\caption{Modified Lindemann parameter $\Gamma_L(t)$ for different values of 
$\dot \gamma$ (as reported in the legend)}
\label{Fig.lindemann} 
\end{figure}

Our results are presented in Fig. \ref{Fig.lindemann}.
In the elastic regime (small $\dot \gamma$), the Lindemann parameter $\Gamma (t)$
exhibits a plateau at ``long'' times confirming the crystalline intralayer structure.
At higher $\dot \gamma$ (i.e., $\dot \gamma \geq 60$) the situation gets more
complicated. For $60/\tau \leq \dot \gamma \leq 100/\tau$, $\Gamma_L (t)$ diverges proving a liquid behavior. 
While this feature was clearly expected for $\dot \gamma = 60/\tau$ and $80/\tau$ from our
{\it static} analysis of $g(r)$ [see Fig. \ref{Fig.g_of_r_shear}(b)], that was not
obvious for  $\dot \gamma = 100/\tau$.
It is therefore only at very high shear rate (i.e., $\dot \gamma \geq 200$) that {\it true}
intralayer crystalline-reordering is recovered, as indicated by the existence of the plateau
in $\Gamma_L (t)$ whose value is comparable to that obtained in the elastic regime.

%
%

\subsection{Relaxation after cessation of shear}
\label{sec:relax-after-cess}

\begin{figure}
\includegraphics[width = 8.5 cm]{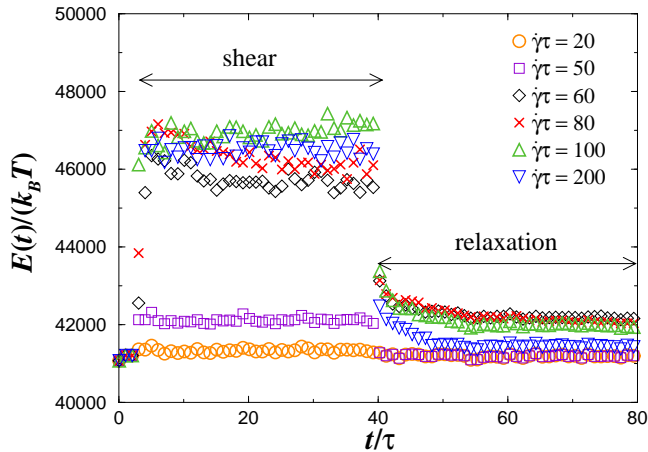}
\caption{Time evolution of the total potential energy of interaction $E(t)$:
before - during - and after shear. The values of $\dot \gamma$, considered 
during the shear process, are reported in the legend.}
\label{Fig.relax_energy} 
\end{figure}

\begin{figure}
\includegraphics[width = 15 cm]{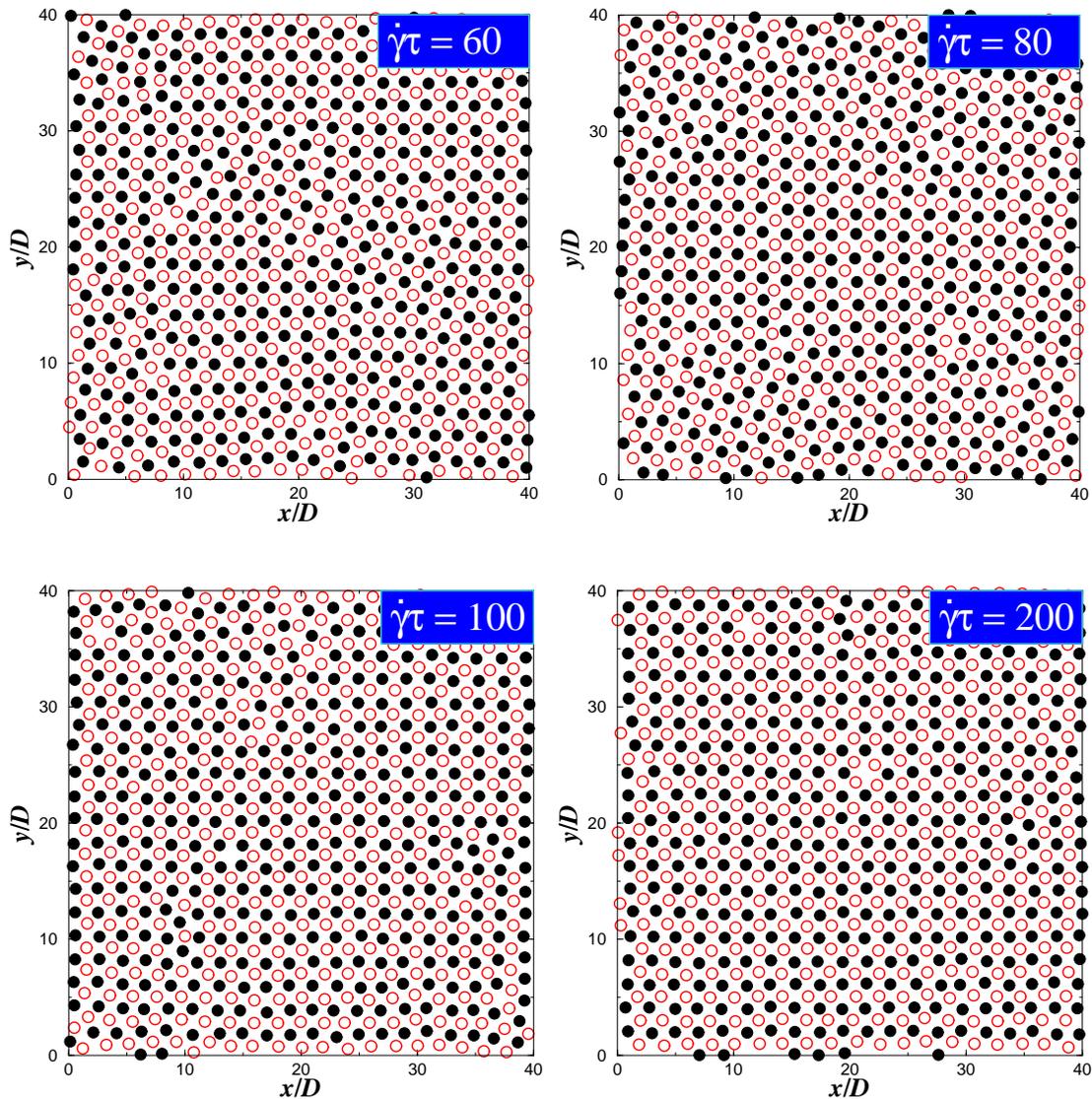}
\caption{Simulation snapshots of relaxed systems taken at $t=80\tau$
for different values of the prior applied shear rates $\dot \gamma$ (as indicated).}
\label{Fig.relax_snap} 
\end{figure}

We now investigate how the system gets back
to equilibrium after cessation of shear.
A suitable and simple way to study a relaxation process is to monitor 
the evolution in time of the total potential energy of interaction
$E(t)=V_{yuk}+V_{wall}$. In our simulations, the cessation of shear occurs at $t=40\tau$.
Profiles of $E(t)$  for different shear rates $\dot \gamma$ applied
{\it prior} relaxation are plotted in Fig. \ref{Fig.relax_energy}.
The corresponding microstructures at long time $t=80\tau$
for $60 \gamma / \tau \leq \dot \gamma \leq 200\gamma / \tau$
are sketched in Fig. \ref{Fig.relax_snap}.
For low $\dot \gamma$ (here $\dot \gamma \leq 50\gamma / \tau$),
the relaxation process is very fast as it should be.
Note that the equilibrium energy value is not exactly 
recovered because of the existence of some long-living defects.

The relaxation process gets qualitatively different for highly
sheared systems (here $\dot \gamma \geq 60\gamma / \tau$).
For the samples that have undergone a shear-induced melting as
deduced from our criterion based  on $\Gamma_L(t)$ 
[see Fig. \ref{Fig.lindemann} with $\dot \gamma \tau = 60,80,100$],
we remark that they all exhibit a similar relaxation behavior
[see Fig. \ref{Fig.relax_energy} with $\dot \gamma \tau = 60,80,100$].
In particular the relaxation is thereby much slower, partly due to the existence of
{\it many} long living defects. Those latter also explain the high energy reported
in the long time scale. There are several defects such as dislocations, 
(low angle) grain boundaries (especially for $\dot \gamma \tau = 60,80$) 
and vacancies that are easily identifiable in the snapshots of 
Fig. \ref{Fig.relax_snap}.

On the other hand at large enough $\dot \gamma$ (here $\dot \gamma = 200/\tau$)
the relaxation is faster as indicated by the faster earlier occurrence of
a $E(t)$-plateau (which is also deeper). Nonetheless, the energy of this (nearly)
relaxed system remains higher than those that were weakly sheared 
($\dot \gamma<\dot \gamma_0$). Again the existence of some vacancies 
(see Fig. \ref{Fig.relax_snap} with $\dot \gamma = 200/\tau$) 
increases the energy system as well as the time of {\it full} relaxation.

%
%

%
%

\section{Conclusions
 \label{Sec.conclu}}

To conclude we perform Brownian dynamics computer simulations to
study crystalline bilayers of charged colloidal suspensions which are confined 
between two parallel plates and sheared via a relative motion of the 
two plates. 
For the parameters under consideration, 
the unsheared equilibrium configuration are two crystalline layers with a 
nested quadratic in-plane structure.
For increasing shear rates $\dot \gamma$, we find the following
steady states: first, there is a static solid which is elastically sheared 
until a yield-stress limit is reached. 
Higher shear rates melt the crystalline bilayers and even higher shear rates lead 
to a reentrant solid stratified in the shear direction.
We have then studied the relaxation of the sheared steady state back to 
equilibrium after an instantaneous cessation of shear and found a nonmonotonic 
behavior of the typical relaxation time as a function of the shear rate $\dot \gamma$.
In particular, application of (very) high shear rates accelerates the (post-)relaxation 
back to equilibrium since shear-ordering facilitates the growth of the equilibrium crystal. 

\acknowledgments 
We thank T. Palberg and C. Wagner for helpful comments. 
This work was supported by the DFG within the Transregio
SFB TR6 (project section D1).


\end{document}